\documentclass[12pt]{article}
\begin{document}
\vspace*{-.6in}
\thispagestyle{empty}
\begin{flushright}
CALT-68-2289\\ 
CITUSC/00-042\\
hep-th/0007117
\end{flushright}
\baselineskip = 18pt

\vspace{.5in}
{\LARGE
\begin{center}
Reminiscences of Collaborations with Jo\"el Scherk\end{center}}
\vspace{.5in}

\begin{center}
John H. Schwarz\footnote{Work supported in
part by the U.S. Dept. of Energy under Grant No.
DE-FG03-92-ER40701.}
\\
\emph{California Institute of Technology\\ Pasadena, CA  91125, USA}
\end{center}
\vspace{1in}

\begin{center}
\textbf{Abstract}
\end{center}
\begin{quotation}
\noindent
I had the privilege of collaborating with Jo\"el Scherk
on three separate occasions:  in 1970 at Princeton, in 1974 at Caltech,
and in 1978-79 at the Ecole Normale Sup\'erieure. In this talk I give
some reminiscences of these collaborations.\end{quotation}

\vfil
\centerline{Presented at \it Conf\'erence anniversaire du LPT-ENS}

\newpage

\pagenumbering{arabic}

\section{Our First Meeting}

In 1969 my duties as an Assistant Professor in Princeton
included advising some assigned graduate students.
The first advisees who came to see me were a pair of Frenchmen,
Andr\'e Neveu and Jo\"el Scherk. I had no advanced warning about them,
and so I presumed they were just another pair of entering students. In fact,
they had achieved the equivalent of a Ph.D. in France and were attending Princeton
on some special Fellowships. Because their degrees were not called PhDs, the
Princeton bureaucracy classified them as graduate students, and so they were
assigned to me.

At our first meeting, I asked the usual questions: ``Do  you need
to take a course on electrodynamics?",  ``Do you need to take a
course on quantum mechanics?'', etc. They assured me that they
already had learned all that, and it wouldn't be necessary. So I
said okay, signed their cards, and they left.

\section{Loop Amplitudes}

Veneziano discovered his famous formula for a four-particle amplitude in 1968
\cite{Veneziano:1968}.
In 1969 various groups constructed $N$-particle generalizations
of the Veneziano amplitude \cite{Bardakci:1969a} - \cite{Koba:1969b} 
and showed that they could be consistently
factorized on a well-defined spectrum of single-particle states as required
for the tree approximation of a 
quantum theory \cite{Fubini:1969a} -
\cite{Fubini:1970}.  In those days the
theory in question was called the dual resonance model. 
Today we would refer to it
as the bosonic string theory. Knowing
the tree approximation spectrum and couplings,
it became possible to construct one-loop
amplitudes. The first such attempt was made by Kikkawa, Sakita,
and Virasoro \cite{Kikkawa:1969}.
They did not have enough information in hand to do
it completely right, but they pioneered many of the key ideas and
pointed the way for their successors.
Around this time (the fall of 1969) I began studying one-loop amplitudes
in collaboration with
David Gross, who was also an Assistant Professor at Princeton.
(We had collaborated previously when we were graduate
students at Berkeley.)

A couple months after our first meeting,  
Jo\"el and Andr\'e reappeared in my office and said
that they had found some results they would like to show me. 
They proceeded to
explain their analysis of the divergence in the planar 
one-loop amplitude.
They realized that by performing a Jacobi transformation of 
the theta
functions in the integrand they could isolate the divergent 
piece and propose
a fairly natural counterterm \cite{Neveu:1970}.  
I was very impressed by this
achievement. It certainly convinced me that they did not need 
to take any
more quantum mechanics courses!
Essentially the same calculation was carried out independently 
and simultaneously
by Susskind and Frye \cite{Frye:1970a}. The modern interpretation of the
result is that viewed in a dual channel there is a closed string 
going into the vacuum.
The divergence can be attributed to the tachyon in that channel, 
and its contribution
is the piece that they subtracted. The same result can also 
be obtained by analytic continuation techniques.
This interpretation explains why in a model without tachyons
such divergences would not occur. The cancellation of 
the milder divergences due
to dilaton tadpoles that can appear in superstring theories
became an important consideration in later years.

Since Neveu and Scherk were working on problems that were closely
related to those that Gross and I were studying, we decided to 
join forces.
One of the important things that the four of us 
discovered \cite{Gross:1970b}
was that the nonplanar loop amplitude
contains not only the expected two-particle threshold 
singularities but also new and
unexpected singularities.
This was also discovered independently by Frye and Susskind at about
the same time \cite{Frye:1970b}, In both of these works the dimension of
spacetime was assumed to be four, and the Virasoro subsidiary
constraints were not implemented on the internal states
circulating in the loop. As a result the singularities were found
to be unitarity-violating branch points.
We wanted to identify the leasing Regge trajectory associated to
these singularities with the Pomeron, since they carried vacuum
quantum numbers, but clearly something wasn't quite right.

About a year later Lovelace observed that if one allows the
spacetime dimension to be 26 and supposes that the subsidiary
conditions imply that only transverse oscillators contribute, then
instead of branch points the singularities would be poles \cite{Lovelace:1971}.
As we now know, these are are
the closed-string poles in the nonplanar open-string loop. This
calculation showed that unitarity requires that one choose the
critical dimension and the intercept value for which the Virasoro
conditions are satisfied.  It was supposed that
the unphysical Regge intercept value of two -- implying the
existence of a massless spin two particle as well as a tachyon --
would somehow be lowered to the desired Pomeron value of one in a more
realistic model.

As I recall, Lovelace's paper was quite a shock to everyone, since
until then nobody considered allowing the dimension of spacetime
to be anything other than four. We were doing hadron  physics,
after all, and four was certainly the right answer. Before long,
most people were convinced that this theory required 26
dimensions. However, it had a tachyon and no fermions, so it was
unphysical anyway. It was natural to suppose that a better theory
would require four dimensions. We did find a better theory soon
thereafter. However, it turned out to require ten dimensions, not
four.

\section{String Theory for Unification}

In 1971 Scherk returned to Paris, and
in 1972 I moved to Caltech. At Caltech,
Murray Gell-Mann
put ample funds at my disposal to invite collaborators of my 
choosing.
One of them was
Jo\"el Scherk, who spent the first half of 1974 visiting Caltech.
I am certainly happy that he did!
 
Prior to coming to Caltech, 
he had visited NYU,
where he had written a very elegant review of string theory \cite{Scherk:1975}.

The hadronic interpretation of string theories was plagued not
only by the occurrence of massless vector particles in the
open-string spectrum, but by a massless tensor particle in the
closed-string spectrum, as well. Several years of effort were
expended on trying to modify each of the two string theories so as
to lower the leading open-string Regge intercept from 1 to 1/2 and
the leading closed-string Regge intercept from 2 to 1, since these
were the values required for the leading meson and Pomeron Regge
trajectories. Some partial successes were achieved, but no fully
consistent scheme was ever developed. Furthermore, 
efforts to
modify
the critical spacetime dimension
from 26 or 10 to four  also led to difficulties.

By
1974, almost everyone who had been working on string theory
had dropped it and moved to greener pastures. The standard model
had been developed, and was working splendidly.
Against this backdrop, Jo\"el and I
(stubbornly) decided to return to the nagging unresolved problems
of string theory. We felt that the theory has such a compelling mathematical structure that
it ought to be good for something.
Before long our focus shifted to the question of
whether the massless spin two particle in the spectrum interacted
at low energies in accordance with the dictates of general
relativity, so that it might be identified as a graviton.
Several years earlier Jo\"el and Andr\'e Neveu had studied the massless
open-string spin one states and showed that in a suitable low-energy limit
they interacted precisely in agreement with
Yang--Mills theory \cite{Scherk:1971, Neveu:1972}.
Now we wondered about the analogous question for the massless spin two closed-string
ground state.
Roughly, what we proved was that the theory has the gauge
invariances required to decouple unphysical polarization states.
Then it followed that the interactions at low energy must be those of general relativity.

Once we had digested the fact that string theory inevitably contains gravity we
were very excited. We knew that string theory does not have
ultraviolet divergences, because the short-distance structure is
smoothed out, but that any field-theoretic approach to gravitation
necessarily gives nonrenormalizable ultraviolet divergences.
Evidently, the way to make a consistent quantum theory of gravity
is to posit that the fundamental entities are strings rather than
point particles \cite{Scherk:1974a}. Adopting this viewpoint meant that the length
scale of string theory had to be identified with the Planck scale
rather than the QCD scale, which represented a change of almost 20
orders of magnitude. So, even though the mathematics was largely
unchanged, this was a large conceptual change.

Convinced of the
importance of this viewpoint, we submitted a short essay summarizing the
argument to the Gravity Research Foundation's 1975 Essay competition \cite{Scherk:1975a}.
Regrettably, the judges were not impressed.
 (We received `Honorable Mention'.)

Having changed the goal of string theory to the problem
of describing quantum gravity rather
than hadronic physics, it became natural to suppose that the
Yang-Mills gauge interactions that string theory also contains should
describe the other forces.  This means that one is dealing with a unified
quantum theory -- an explicit realization of Einstein's dream.
(Actually, it's not clear that Einstein wanted his unified theory
to be quantum, but that's another story.) Moreover, the existence
of extra dimensions could now be a blessing rather than a curse.
After all, in a gravity theory the geometry of spacetime is
determined dynamically, and one could imagine that the extra
dimensions form some kind of compact space 
(spontaneous
compactification).
We attempted to construct a specific compactification scenario
in a subsequent paper \cite{Scherk:1975b}.
From today's vantage point, that work looks rather primitive.

In Japan, Yoneya independently realized that the massless spin two
state of string theory interacts in accordance with the dictates
of general relativity \cite{Yoneya:1974}.
I would claim, however, that Scherk and I were only
ones to take the next step and to propose seriously that string
theory should be the basis for constructing a unified quantum
theory of all forces. In any case, the recognition of that possibility represented a
turning point in my research. I found the case convincing and was
committed to exploring the ramifications,
which is what I have been doing ever since.

I still do not
understand why it took another decade until a large segment of the
theoretical physics world became convinced that string theory
was the right approach to unification. (There were a few
people who caught on earlier, of course -- most notably Lars Brink, David Olive, and
Michael Green. I also received encouragement throughout this
period from Murray Gell-Mann. By the early 1980s Edward Witten and
Bruno Zumino were also very supportive.)
One of my greatest regrets
is that Jo\"el was not alive to witness the impact that this idea would
eventually have.

\section{Spacetime Supersymmetry}

Supersymmetry arose in string theory and in field theory
separately. However, for the first several years, the
supersymmetry considered by string theorists only pertained to the
two-dimensional world-sheet theory. I will not review that story
here, since I will be speaking about it at a conference in
Minnesota in October. In any case, Jo\"el and I became interested in
supersymmetry theories, and we each did some work on supersymmetric
field theories. One work by the two of us and Lars  Brink was the first 
construction of supersymmetric Yang-Mills theories in various dimensions
\cite{Brink:1977}. When this work was done, Brink and I were at
Caltech and Scherk was in Paris. We communicated by mail. (Email
was not an option.) We found that the requisite gamma matrix
identity required by these theories $\gamma^m_{(ab}\gamma^m_{ c)d}
=0$ could be satisfied in dimensions 3, 4, 6, and 10.  (The fact
that the number of transverse dimensions is 1, 2, 4, or 8 suggests
a connection with the number fields: real, complex, quaternionic,
octonionic. However, the precise way this should work is not clear
even today.)

At about the same time as the work by Brink, Scherk, and me,
Scherk also did some very important
work with Gliozzi and Olive \cite{Gliozzi:1976, Gliozzi:1977}. It gave further
explanation of the results about super Yang--Mills theory, and took the next major
step. This was to conjecture a connection with the RNS string theory.
Specifically, they noted that if the spectrum of the 10-dimensional theory
was restricted in a specific consistent way (GSO projection) that then the
number of bosons and fermions at each mass level would be exactly the same.
Moreover, since the spectrum contains a massless gravitino in the closed string sector
this truncation is essential for consistency.
This constituted powerful evidence (though not a proof) that the GSO-projected
RNS string has local 10-dimensional spacetime supersymmetry. I was very
delighted by this result. Finally we had a theory that was tachyon-free and
seemed to make sense as a starting point for a unified theory. I was committed
to exploring it more deeply.

There was much that needed to be learned before it would be possible
to really dig into superstring theory.  For example, one needed to understand supergravity,
which was just starting to be developed. One of the most beautiful results
ever found in supergravity was the action of
11-dimensional supergravity by Cremmer, Julia, and Scherk \cite{Cremmer:1978}.
That said, I must admit that when it first appeared I was a bit baffled.
My problem was that I knew that supergravity could not give a consistent
quantum theory by itself. So I felt that the only supergravity theories
worth studying were those that might arise from string theories at
low energy. However, since superstring theory only allowed ten dimensions,
this did not seem to leave a role for 11-dimensional supergravity.
I viewed it as a 10\% error. As we now
know, what I failed to consider was the possibility that in (Type IIA) superstring
theory an 11th dimension
might be generated nonperturbatively!

\section{Supersymmetry Breaking}

I spent the Academic Year 1978--79 visiting the Ecole Normale, on leave of absence from
Caltech. I was eager to work with Scherk on supergravity, supersymmetrical strings,
and related matters. He was struggling with rather serious health problems
during that year, so he wasn't able to participate as fully as when we were in Caltech
five years earlier,
but he was able to work about half time. On that basis we were able to
collaborate successfully and to obtain some results
that I think are interesting.

After various wide-ranging discussions we decided
to focus on the problem of supersymmetry breaking. We wondered how, starting
from a supersymmetric string theory in ten dimensions, one could end up with
a nonsupersymmetric world in four dimensions. The specific supersymmetry breaking mechanism that we discovered can be explained classically
and does not really require strings, so we explored it in
a field theoretic setting. The idea is that in a theory with extra dimensions and
global symmetries that do not commute with supersymmetry ($R$ symmetries
and $(-1)^F$ are examples),
one could arrange for a twisted compactification, and that
this would break supersymmetry. For example, if one extra dimension
forms a circle, the fields when continued around the circle could come
could back transformed by an R-symmetry group element. If the gravitino,
in particular, is transformed then it acquires mass in a consistent manner.
We called this type of local supersymmetry breaking ``spontaneous". Whether
this is an appropriate use of that term can be debated. The important thing is that
the symmetry breaking is consistent and sensible.

An interesting example of our supersymmetry breaking mechanism was
worked out in a paper we wrote together with Eugen Cremmer \cite{Cremmer:1979}.
We started with maximal supergravity in five dimensions. This theory
contains eight gravitinos that transform in the fundamental representation
of a USp(8) R-symmetry group. We took one dimension to form a circle of radius $R$,
and examined the resulting four-dimensional theory keeping the lowest
Kaluza--Klein modes. The supersymmetry-breaking R-symmetry element is
a USp(8) element that is characterized by four real mass parameters, since this
group has rank four. These four masses give the masses of the four complex
gravitinos of the resulting four-dimensional theory. In this way we were able
to find a consistent four-parameter deformation of ${\cal N} = 8$ supergravity.
(Most good ideas from previous years have reappeared, sometimes in new guises,
in modern superstring theory. This construction is an example of one that
has not reappeared yet, so I wonder if there may be an opportunity here.)

Even though the work that Scherk and I did on supersymmetry breaking was motivated
by string theory, we only discussed field theory applications in our articles. The
reason I never wrote about string theory applications was that in the string theory
setting I could not see how to decouple the supersymmetry breaking mass parameters
from the compactification scales.  This was viewed as a serious problem because
the two scales are supposed to be hierarchically different. In recent times, people
have been considering string theory brane-world scenarios  in which much larger
compactification scales are considered. In such a context our supersymmetry breaking
mechanism might have a role to play. Indeed, quite a few authors have explored
various such possibilities in recent times.

\section{Conclusion}

When I left Paris in the summer of 1979 I visited CERN for a
month. There I began a collaboration with Michael Green. During
that month we began to formulate a plan for exploring how the
spacetime supersymmetry identified by Gliozzi, Scherk, and Olive
is realized in the interacting theory. However, it took almost a
year for us to achieve definitive results. Therefore, in the fall
of 1979,  when I spoke at a conference on supergravity that was
held in Stony Brook, I reported on the work that Scherk and I had
done during the preceding year \cite{Schwarz:1979}.  At the same
conference, J\"oel gave a talk entitled {\it From Supergravity to
Antigravity} \cite{Scherk:1979c}. He was intrigued by the fact
that in string theory graviton exchanges are accompanied by
antisymmetric tensor and scalar exchanges that can cancel the
gravitational attraction. I felt that the use of the term {\it
antigravity} was a bit too sensational. In any case, nowadays we
understand that the effect he was discussing is quite important.
For example,  parallel BPS D-branes form stable supersymmetric
systems precisely because the various forces cancel.

The Stony Brook conference was the last time that I saw Jo\"el. I
was very shocked and saddened a few months later when I was informed that
he had passed away. As you know, the ideas that he had helped to pioneer
have been very influential in the subsequent development of our field.
It is a pity that he was not able to participate in these developments
and to enjoy the recognition that he would have received.

\end{document}